\documentclass[epj]{svjour}
\usepackage{graphics}
\usepackage{graphicx}
\usepackage{epstopdf}
\usepackage{amssymb}   
\usepackage{url}
\usepackage{amsmath}%vbs added 
\begin{document}
\title{Advances and new ideas for neutron-capture astrophysics experiments at CERN n\_TOF}

\author{%
C.~Domingo-Pardo\inst{1,*}, 
V.~Babiano-Suarez\inst{1}, 
J.~Balibrea-Correa\inst{1}, 
L.~Caballero\inst{1}, 
I.~Ladarescu\inst{1}, 
J.~Lerendegui-Marco\inst{1}, 
J.L.~Tain\inst{1}, 
A.~Tarife\~{n}o-Saldivia\inst{1}, 
O.~Aberle\inst{2}, 
V.~Alcayne\inst{3}
% ...
%remove \and on previous line
The n\_TOF Collaboration

}
\institute{
$^{1}$Instituto de F\'{\i}sica Corpuscular, CSIC - Universidad de Valencia, Spain;  
$^{2}$European Organization for Nuclear Research (CERN), Switzerland; 
$^{3}$Centro de Investigaciones Energ\'{e}ticas Medioambientales y Tecnol\'{o}gicas (CIEMAT), Spain;  
$^{4}$Laboratori Nazionali di Pavia, Italy;  
$^{5}$Istituto Nazionali di Fisica Nucleare (INFN), Sezione di Catania, Italy;  
$^{6}$University of Lodz, Poland;  
$^{7}$University of Manchester, United Kingdom;  
$^{8}$CEA Irfu, Universit\'{e} Paris-Saclay, France;  
$^{9}$Horia Hulubei National Institute of Physics and Nuclear Engineering, Romania;  
$^{10}$Department of Physics, Faculty of Science, University of Zagreb, Zagreb, Croatia;  
$^{11}$University of Santiago de Compostela, Spain;  
$^{12}$Universitat Polit\`{e}cnica de Catalunya, Spain;  
$^{13}$Istituto Nazionale di Fisica Nucleare, Sezione di Trieste, Italy;  
$^{14}$Osservatorio Astronomico di Trieste, Italy;  
$^{15}$National Technical University of Athens, Greece;  
$^{16}$Laboratori Nazionali di Torino, Italy;  
$^{17}$Istituto Nazionale di Fisica Nucleare, Sezione di Bari, Italy;  
$^{18}$Istituto Nazionale di Fisica Nucleare, Sezione di Bologna, Italy;  
$^{19}$Universidad de Sevilla, Spain;  
$^{20}$Istituto Nazionale di Fisica Nucleare, Sezione di Perugia, Italy;  
$^{21}$Istituto Nazionale di Astrofisica - Osservatorio Astronomico di Teramo, Italy;  
$^{22}$Goethe University Frankfurt, Germany;  
$^{23}$Instituto Superior T\'{e}cnico, Lisbon, Portugal;  
$^{24}$Physikalisch-Technische Bundesanstalt (PTB), Braunschweig, Germany;  
$^{25}$Paul Scherrer Institut (PSI), Villigen, Switzerland;  
$^{26}$University of Ioannina, Greece;  
$^{27}$Agenzia nazionale per le nuove tecnologie (ENEA), Bologna, Italy;  
$^{28}$Joint Institute for Nuclear Research (JINR), Dubna, Russia;  
$^{29}$University of Granada, Spain;  
$^{30}$Laboratori Nazionali di Frascati, Italy;  
$^{31}$European Commission, Joint Research Centre (JRC), Geel, Belgium;  
$^{32}$University of York, United Kingdom;  
$^{33}$TU Wien, Atominstitut, Wien, Austria;  
$^{34}$Helmholtz-Zentrum Dresden-Rossendorf, Germany;  
$^{35}$Japan Atomic Energy Agency (JAEA), Tokai-Mura, Japan;  
$^{36}$Charles University, Prague, Czech Republic;  
$^{37}$School of Physics and Astronomy, University of Edinburgh, United Kingdom;  
$^{38}$Dipartimento di Fisica e Astronomia, Universit\`{a} di Bologna, Italy;  
$^{39}$Istituto Nazionale di Fisica Nucleare, Sezione di Legnaro, Italy;  
$^{40}$Dipartimento Interateneo di Fisica, Universit\`{a} degli Studi di Bari, Italy;  
$^{41}$Consiglio Nazionale delle Ricerche, Bari, Italy;  
$^{42}$Uppsala University, Sweden;  
$^{43}$Institut Laue Langevin, France;  
$^{44}$Dipartimento di Fisica e Astronomia, Universit\`{a} di Catania, Italy ; 
$^{45}$Leibniz-Institut f\"ur Festk\"orper- und Werkstof\/f\/forschung Dresden (IFW) e.V., Germany
%remove \and on previous line
}

%\author{First author\inst{1} \and Second author\inst{2}% etc

%}                     % Do not remove
%\institute{Instituto de F{\'\i}sica Corpuscular, CSIC-University of Valencia, Valencia, Spain}

\date{Received: \today / Revised version: \today}

\abstract{
This article presents a few selected developments and future ideas related to the measurement of $(n,\gamma)$ data of astrophysical interest at CERN n\_TOF. The MC-aided analysis methodology for the use of low-efficiency radiation detectors in time-of-flight neutron-capture measurements is discussed, with particular emphasis on the systematic accuracy. Several recent instrumental advances are also presented, such as the development of total-energy detectors with $\gamma$-ray imaging capability for background suppression, and the development of an array of small-volume organic scintillators aimed at exploiting the high instantaneous neutron-flux of EAR2. Finally, astrophysics prospects related to the intermediate $i$ neutron-capture process of nucleosynthesis are discussed in the context of the new NEAR activation area.
}
\PACS{
      {neutron capture} {} \and
      {time-of-flight} {} \and 
      {s-process} {} \and 
      {i-process} {} \and
      {nucleosynthesis}{}
     } % end of PACS codes
 %end of abstract
%
\maketitle
\section{Introduction}\label{sec:introduction}

\begin{figure*}[!hbtp]
    \centering
    \includegraphics[width=2\columnwidth]{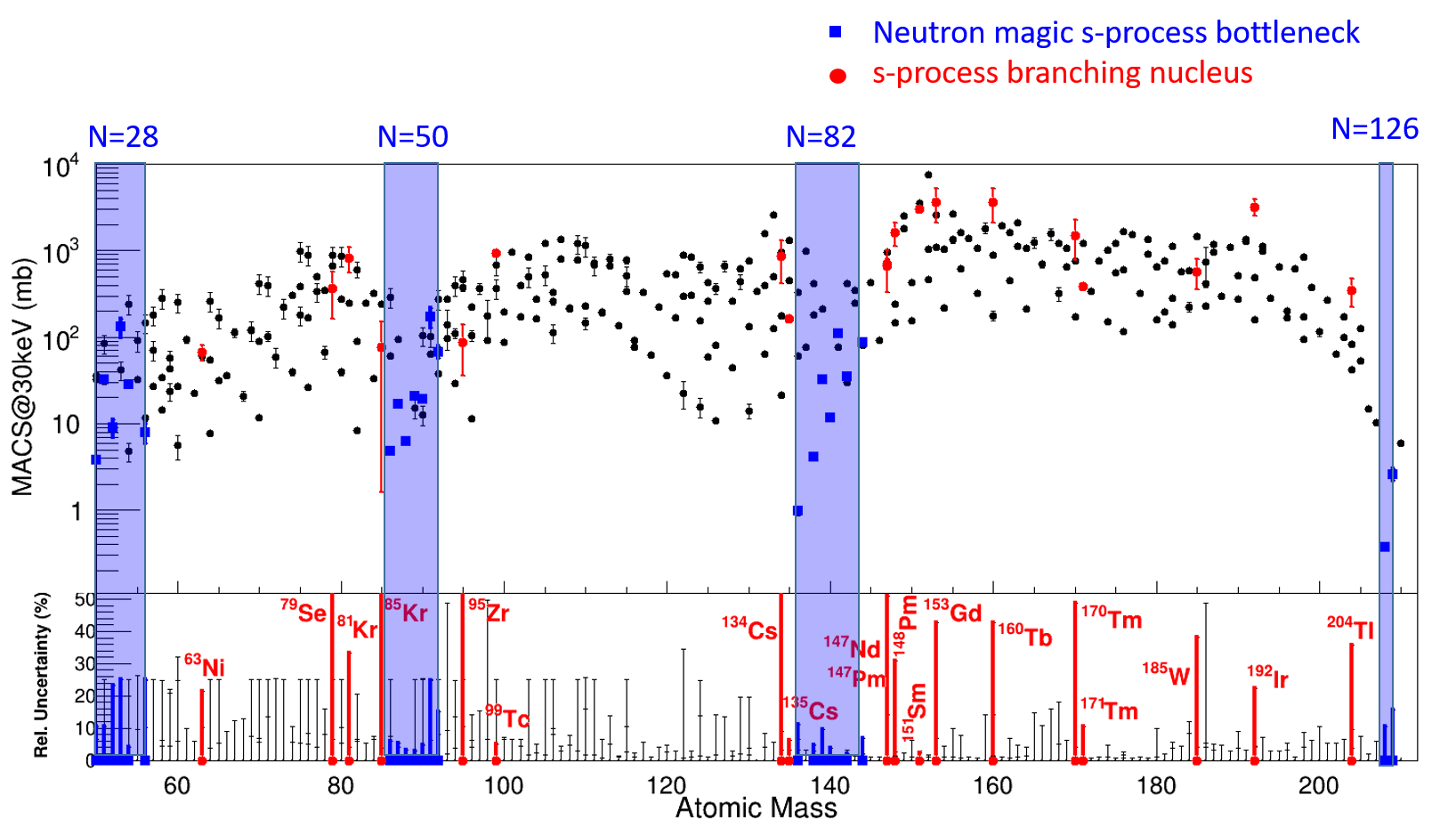}
    \label{fig:macs}
    \caption{Maxwellian averaged neutron-capture cross section at $kT=30$~keV (top panel) and their relative uncertainties (bottom panel). Blue colors refer to nuclei with neutron-shell closures and branching nuclei are displayed in red. Most values are taken from \cite{kadonisv1} (see text for details).}
\end{figure*}

The fundamental role of neutron-induced reactions in the formation of the heavy elements in the universe was already evident in 1948~\cite{ABG48,Alpher48a,Alpher48b,Alpher48c}, although it was probably the first observation of technetium in S-type stars~\cite{Merrill52} and the subsequent quantitative theory of nucleosynthesis~\cite{BBFH,Cameron57}, which triggered and guided an enormous experimental effort, that still prevails today~\cite{Gibbons67,Reifarth05,Kaeppeler11,Langanke18,Estrade19,Massimi22,Schatz22}. 
%To quote one illustrative example, low-efficiency Moxon-Rae detectors were invented to bridge the gap in detection sensitivity with respect to former large-scintillation tanks. This advance was sufficient to access for the first time neutron-capture rates on isotopically enriched samples of tin-nuclei. This development revealed for the first time the inverse-proportionality between neutron-capture cross section at $\sim$25~keV and isotopic s-process abundances, which represented one of the first experimental confirmations of the s-process theory.
This article describes some experimental developments primarily aimed at measuring nuclear data of interest for nucleosynthesis studies in hydrostatic stages of stellar evolution, namely asymptotic giant branch (AGB-) and massive-stars~\cite{Kaeppeler11,Pignatari10}. These works were carried out at the n\_TOF facility, which has been extensively described in detail elsewhere~\cite{Colonna18,Chiaveri20,Esposito21}. The first topic reported in Sec.~\ref{sec:phwt} is related to the accuracy of the measurements carried out in neutron time-of-flight (TOF) experiments using low-efficiency radiation detectors. This is an important subject for astrophysics because data from many previous measurements still exhibit cross-section uncertainties that are significantly larger than the few percent uncertainty attainable from stellar observations or meteorites analysis\cite{Kaeppeler11}. The experimental situation is illustrated in Fig.\ref{fig:macs}, which shows Maxwellian average cross sections (MACS) at $kT=30~$~keV and current uncertainties~\cite{kadonisv1} for all nuclei involved in $s$-process nucleosynthesis. As pointed out in several recent sensitivity studies~\cite{Neyskens15,Cescutti18,Cescutti19,Nishimura18a,Nishimura18b}, the cross sections of many isotopes need to be re-measured either with improved accuracy or over more complete neutron-energy ranges in order to derive reliable information of astrophysical interest. This is particularly true for the seeds of the $s$ process around the Fe-Ni region~\cite{Cescutti18}, whose cross sections at $kT=30$~keV still show relatively large uncertainties (see bottom panel in Fig.~\ref{fig:macs}). Following this logic, many neutron-capture experiments were made at n\_TOF over the last 20~years~\cite{Massimi22} and many more experiments on stable isotopes will follow in the coming years. The new measurements will benefit, not only from the enhanced accuracy approach described below in Sec.~\ref{sec:phwt}, but also from new instrumental developments such as those reported in Sec.~\ref{sec:iTED} and Sec.~\ref{sec:sTED}.

Another topic which focuses many experimental efforts nowadays is the determination of  neutron-capture cross sections on unstable nuclei~\cite{Kaeppeler11}. In AGB- and massive stars, radioactive nuclei may split the nucleosynthesis path and yield a local isotopic pattern around the branching nucleus, which is very sensitive to the physical conditions of the stellar environment. Therefore, neutron-capture measurements of these nuclei provide stringent constraints on stellar structure and evolution models. As shown in Fig.~\ref{fig:macs} several $s$-process branching nuclei have been measured with high accuracy~\cite{Kaeppeler11,Massimi22}, but there is still a significant number of them that have not been accessed yet owing to limitations in state-of-the-art detection systems and sample-production capabilities. Sec.~\ref{sec:iTED} and Sec.~\ref{sec:sTED} describe some recent technical developments aimed at enhancing detection sensitivity in neutron-capture experiments, either by means of $\gamma$-imaging or by means of very-low efficiency detectors. Sec.~\ref{sec:cycling} then describes new ideas at n\_TOF intended to afford direct neutron-capture measurements of interest for more exotic stellar environments, such as the intermediate $i$-process of nucleosynthesis~\cite{Cowan77}. Finally, Sec.~\ref{sec:summary} summarizes the main conclusions and future prospects.

\section{Improved accuracy measurements via MC-aided PHWT}\label{sec:phwt}
One of the most relevant aspects when dealing with experimental data concerns the systematic accuracy of the measurement, the proper identification of experimental uncertainties and their realistic assessment. Therefore, in the first n\_TOF experimental campaign in 2001 a study~\cite{Tain04} was carried out in order to address the systematic accuracy attainable with the so-called Pulse-Height Weighting Technique (PHWT). Originally developed in the sixties at ORNL in a pioneer work by R.L.~Macklin et al.~\cite{Macklin67}, the PHWT has been extremely helpful and very extensively used at different laboratories worldwide for the determination of neutron-capture data of astrophysical interest~\cite{Gibbons67,Kaeppeler11}. The Total-Energy Detection (TED) principle in combination with the PHWT allowed one to virtually mimic an ideal Moxon-Rae detector~\cite{Moxon63}. However, the new approach was much more flexible in terms of apparatus and permitted to attain higher efficiency and better detection sensitivity~\cite{Gibbons67}. The latter was a key aspect to access neutron-capture reactions of astrophysical relevance~\cite{Macklin67}, including also radioactive isotopes such as the $s$-process branchings $^{93}$Zr~\cite{Macklin85} and $^{99}$Tc~\cite{Winters87}. 

An interesting aspect of the TED principle applied with the PHWT is the fact that, essentially, only the requirement of using low-efficiency $\gamma$-ray detectors needs to be experimentally fulfilled~\cite{Macklin67}. 
This opens up a wide scope of options in terms of instrumentation, an aspect that has been also explored and exploited at n\_TOF during the last years, as described later in Sec.~\ref{sec:iTED} and Sec.~\ref{sec:sTED}. Obviously, other additional conditions are required for neutron-TOF experiments, such as fast time-response and low intrinsic sensitivity to scattered neutrons. 
%The proportionality condition between $\gamma$-ray energy and efficiency required also by TED systems can be achieved today practically for almost any detector by means of suitable MC-calculations~\cite{Tain04,Borella07}, as it is also discussed later in Sec.~\ref{sec:iTED}.

However, for several decades the systematic accuracy attainable with the PHWT was a topic of controversy and debate. As clearly stated by F.~Corvi~\cite{Corvi88}, one of the most puzzling aspects in the eighties was a 20\% discrepancy between capture- and transmission-measurements found for the 1.15~keV resonance in $^{56}$Fe($n,\gamma$). At that time, this was quoted as "one of the four major outstanding neutron data problems in the field of fission reactor neutronics"~\cite{Coates83}. The 1.15~keV resonance in $^{56}$Fe represents indeed an ideal case for testing the accuracy of the technique because the capture TOF experiment is mainly sensitive to the neutron width $\Gamma_{n}$, which is accurately known from transmission measurements~\cite{Perey86}. 

In order to tackle this challenge and eventually develop a general and reliable methodology for the analysis of capture data with the PHWT, at n\_TOF we carried out a detailed Monte-Carlo study~\cite{Tain02} followed by a series of systematic measurements~\cite{Tain04}. The latter involved the use of two different C$_6$D$_6$ detectors and iron samples of three different thicknesses (from 0.5~mm to 2~mm). The general conclusions of this work were essentially two. First, it was understood that the only reliable methodology to apply the PHWT accurately was by means of detailed and realistic Monte Carlo (MC) simulations of the experimental set-up for the determination of the weighting function (WF), which included also a specific simulation for every particular sample used in the capture experiments. Thus, at variance with the original approach~\cite{Macklin67} and later works~\cite{Yamamuro76,Corvi88}, there is not such a thing like "The weighting function of the C$_6$F$_6$ scintillation detector"~\cite{Yamamuro76,Mizuno99,Katabuchi17} or a unique "experimental WF"~\cite{Corvi88}. Instead, a WF needs to be calculated for each capture set-up and for each specific sample measured in the TOF experiment~\cite{Tain04}. For relatively thick samples a resonance-dependent WF may be needed in order to account for the different $\gamma$-ray emission and absorption profiles across the sample thickness~\cite{Borella07}. This effect was relevant, for example, in the measurement of $^{197}$Au($n,\gamma$)~\cite{Massimi10} or $^{232}$Th($n,\gamma$)~\cite{Gunsing12}. Self-shielding effects can also play an important role for some samples or resonances and, therefore, the methodology developed in Ref.~\cite{Borella07} has been included in the R-matrix analysis code REFIT~\cite{refit}. For a recent review on the analysis techniques for neutron induced reaction cross-section data the reader is referred to Ref.~\cite{Schillebeeckx12}.

The fact that the WF and the PHWT accuracy is so dependent on so many experimental details reflects also the level of sensitivity in these measurements, where small changes in the experimental conditions can be quickly reflected in the acquired capture data. In some sense, the new MC-aided approach represented a change of paradigm in the analysis of neutron-capture data using the PHWT, which has been adopted by the scientific community~\cite{Borella07,Ren20}. It is worth to emphasize that the work reported in \cite{Corvi88} and references therein, although did not provide a final solution to this problem, it had a crucial relevance towards understanding its origin. It is worth recalling also that MC simulations using the EGS-transport code were applied in ORNL already in 1988~\cite{Perey88}. However,  the latter work still proposed a single WF for all capture experiments regardless of the sample characteristics.

The second aspect found in \cite{Tain04} to be of relevance for the accuracy of the PHWT was related to the signature of nuclear-structure effects in the response functions measured with the C$_6$D$_6$ detectors. In general, differences are found between the capture-cascade spectrum of the sample under study, and the one used as reference, commonly $^{197}$Au($n,\gamma$). The methodology proposed in Ref.~\cite{Tain04} to account for this effect involves the MC simulation of the full capture cascade for both studied- and reference-samples, and then determine a yield correction factor. Because of the interplay with the nuclear-structure effects, the correction factor may even change from one capture-resonance to another, depending on the level spin and parity~\cite{Domingo06a,Domingo06b,Domingo07a}. The main contributions to the yield-correction factor arise from the different number of counts missing under the detection threshold (typically 150-200~keV), $\gamma$-ray summing effects, angular-distribution effects~\cite{Domingo06b,Domingo07b}, conversion-electrons and, if present, isomeric-states~\cite{Domingo06a}. References quoted represent examples, where such correction factors were crucial to keep the systematic uncertainty within the level of 2-3\% RMS. Finally, this result also highlights the relevance of suitable computing codes and libraries~\cite{Agostinelli03}, methods and models~\cite{Becvar98,Tain07,Valenta17,Moreno22} for simulating the cascade of prompt $\gamma$-rays in neutron-capture experiments.

\section{Background suppression via $\gamma$-ray imaging}\label{sec:iTED}
As discussed in the preceding section, one of the most striking features of the TED principle is related to its versatility, namely enabling the use of almost any detection system with efficiency low enough to satisfy
\begin{equation}
    \varepsilon^{c} = 1- \prod^{N}_{j=1} (1 - \varepsilon^{\gamma}_j) \simeq \sum^N_{j=1} \varepsilon_j^{\gamma}.
\end{equation}
Here, $N$ is the number of emitted $\gamma$-rays, $\varepsilon^c$ represents the capture-detection probability and $\varepsilon^{\gamma}$ the $\gamma$-ray detection efficiency. In addition, the efficiency-energy proportionality, $\varepsilon^{\gamma}_j \propto E^{\gamma}_j$, required to attain the total cascade-energy response $\varepsilon^{c} \propto E^c$ can be achieved by means of the PHWT~\cite{Macklin67}. As mentioned before, the detector response function needs to be also suitable for neutron-TOF experiments. Aiming at reducing neutron-induced backgrounds in the detector itself, organic C$_6$F$_6$ detectors were used in the first experiments~\cite{Macklin67,Corvi88}, which were later replaced by C$_6$D$_6$ further optimized by means of C-fiber encapsulations and other improvements~\cite{Plag03,Tain04}. 

\begin{figure}[!htbp]
    \centering
    \includegraphics[width=\columnwidth]{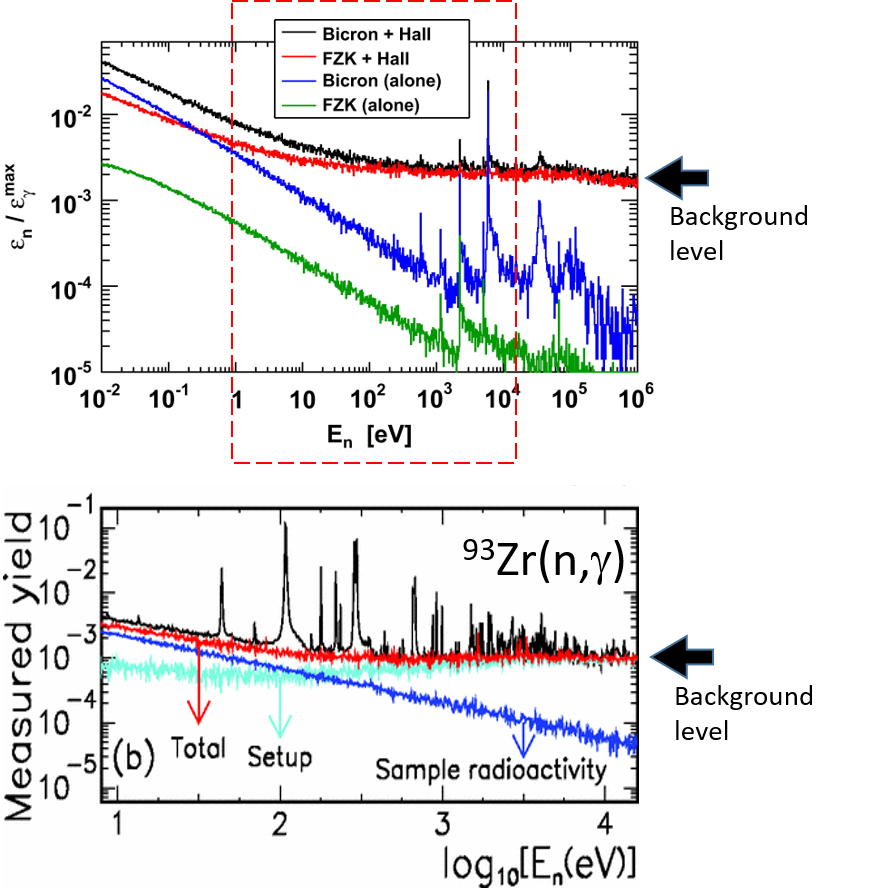}
    \caption{(Top panel) MC simulation~\cite{Zugec14} of the neutron sensitivity, which shows the C$_6$D$_6$-response to neutron-induced $\gamma$-ray background in the walls of the experimental hall. In practice, the resonant structure in the 1~keV-100~keV neutron-energy range is suppressed due to the loss of time-energy correlations for the scattered neutrons (see Ref.~\cite{Zugec16} for details). (Bottom panel) Capture yield of $^{93}$Zr($n,\gamma$)~\cite{Tagliente13}, which shows the limiting effect of the background level in the keV neutron-energy range.}
    \label{fig:zugec}
\end{figure}

Apart from organic scintillation detectors, a NaI(Tl) spectrometer has been used at ANNRI J-PARC~\cite{Mizuno99,Igashira09,Katabuchi14,Katabuchi17}, which actually demonstrates that it is possible to extend the TED principle to very different types of detection systems.
Exploiting further this aspect, a new approach has been investigated at n\_TOF, which applies $\gamma$-ray imaging techniques to discriminate spatially localized $\gamma$-ray background sources~\cite{Domingo16}. This concept seems particularly interesting for the measurement of samples with a small neutron-capture cross section, where neutrons scattered in the sample and subsequently captured in the walls of the experimental hall dominate the background level, instead of neutrons captured directly in the detectors themselves. This situation is depicted in Fig.\ref{fig:zugec}-top, which shows that in the keV neutron-energy region of astrophysical interest the background may be rather dominated by neutrons captured in the walls of the experimental hall, rather than in the detectors themselves~\cite{Zugec14}. The impact of this background is illustrated in Fig.~\ref{fig:zugec}-bottom with the measurement of $^{93}$Zr($n,\gamma$) performed at n\_TOF~\cite{Tagliente13}. As indicated in Ref.~\cite{Neyskens15}, improving the cross-section measurement could help to constrain even more the thermal conditions in AGB stars.
First attempts to apply $\gamma$-ray imaging techniques for background suppression in neutron-capture experiments at n\_TOF employed a pin-hole $\gamma$-camera with a bulky lead collimator attached to a position-sensitive radiation detector~\cite{Perez16}. This work actually demonstrated for the first time the possibility to incorporate imaging techniques in neutron-TOF experiments, although improvements were rather limited owing to the additional background induced by neutrons in the massive collimator itself. The problems ascribed to the use of a massive collimator could be fully overcome by means of an alternative technique based on electronic collimation, originally developed for $\gamma$-ray astronomy~\cite{Schoenfelder73,Wilderman98}. This new approach based on the Compton imaging technique~\cite{Domingo16} has been developed in the framework of the ERC-project HYMNS~\cite{hymns} during the last years at CERN n\_TOF. Compton imaging is based on the use of two or more planes of radiation detectors with both energy- and position-sensitivity operated in time-coincidence. In this way, when a $\gamma$-ray undergoes interaction in several detectors the Compton scattering law can be applied in order to infer information on the incoming radiation direction.
Several technical developments were necessary in order to adapt existing technologies to the field of neutron-capture measurements. These developments were mainly related to the need of achieving good enough energy resolution with SiPMs and large monolithic crystals~\cite{Olleros18}, high spatial resolution and linearity that are challenging due to the big size of the scintillation crystals~\cite{Babiano19,Balibrea21} and implementing a customized dynamic electronic-collimation method for enhanced performance in the Compton imaging~\cite{Babiano20}.

Proof-of-principle experiments~\cite{Babiano21} have been performed at n\_TOF with a prototype of a Total-Energy Detector with imaging capability, called i-TED. These measurements show a significant background reduction in the keV neutron-energy range of interest for astrophysics, when compared to state-of-the-art C$_6$D$_6$ detectors.

\begin{figure}[!htbp]
    \centering
    \includegraphics[width=\columnwidth]{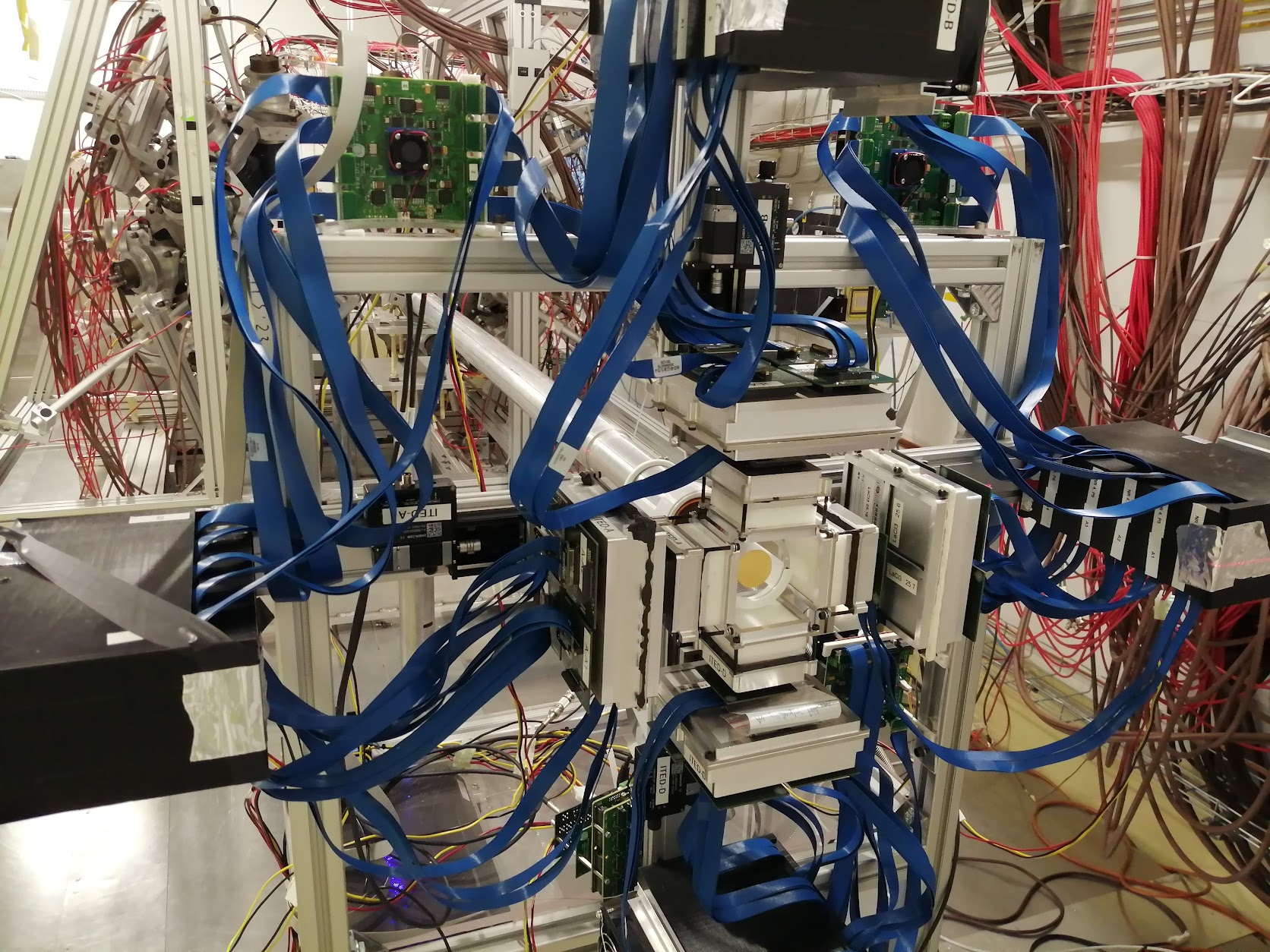}
    \caption{Photograph of the i-TED array during a calibration measurement for the $^{79}$Se($n,\gamma$) experiment in 2022 at CERN n\_TOF EAR1.}
    \label{fig:iTED}
\end{figure}

Fig.~\ref{fig:iTED} shows a picture of the final i-TED system for ($n,\gamma$) experiments, which consists of an array of four large-solid angle Compton cameras in a close configuration around the capture sample. Every Compton module comprises 5 inorganic scintillation crystals, each of them with a size of 50$\times$50~mm$^2$. The front scatter position-sensitive detector has a thickness of 15~mm, whereas the four crystals in the rear absorber plane have a thickness of 25~mm. The modules have been designed in order to maximize detection efficiency, while minimizing neutron-sensitivity in the detectors themselves. To accomplish the latter goal LaCl$_3$(Ce) was preferred versus other options, owing to the relatively small integral capture cross section of Chlorine, and the small contribution of resonances in the keV-energy range of relevance. The Compton modules are supplemented with $^{6}$Li neutron-absorber pads of 20~mm thicknes for reducing further the intrinsic neutron sensitivity of the array (see Fig.~\ref{fig:iTED}).
Pixelated silicon photomultipliers (SiPMs) are used for the readout of the 20 inorganic crystals, leading to a total number of 1280 readout channels. To cope with this large number a dedicated acquisition system based on ASIC TOFPET2 modules~\cite{petsys} was implemented and adapted to this type of experiments. For further details the reader is referred to Refs.~\cite{Babiano20,Babiano21} and references therein.

The i-TED array has been recently applied for the first measurement of the $^{79}$Se($n,\gamma$) capture cross section at n\_TOF EAR1~\cite{Lerendegui21}. A $^{79}$Se sample was produced by high-fluence neutron irradiation in the V4 beam tube of the ILL reactor in Grenoble. To this aim, an eutectic PbSe-alloy sample was prepared at the Paul Scherrer Institut (PSI) in Switzerland, which allowed one to overcome the difficulties ascribed to the low melting point of selenium~\cite{Chiera22}. The measurement with the i-TED array in EAR1 was intended to reduce the large scattered-neutron background arising from the large lead content in the sample, 2.8~g. Furhter, the final PbSe sample had an activity of 5~MBq of $^{75}$Se and 1.6~MBq of $^{60}$Co. Therefore, this sample was also measured at the EAR2 station with the set-up described in the following section.
$^{79}$Se is an important $s$-process branching nucleus, which is particularly well suited to constrain the thermal conditions of the $s$-process in the weak $s$ process~\cite{Walter86,Kaeppeler11,Cescutti19}. Once fully analyzed, the results of this experiment will help to constrain the thermal conditions during core He-burning and shell C-burning in massive stars.

\section{Small-volume C$_6$D$_6$ detectors with high count-rate capability}\label{sec:sTED}
In situations where the background in the experiment is dominated by the decay radioactivity of the sample itself it may become more convenient to exploit the high instantaneous neutron-flux of the EAR2 measuring station. In this way, the relative contribution of the sample radioactivity is minimized with respect to the radiative-capture channel of interest. 
The large instantaneous neutron-flux of n\_TOF EAR2~\cite{Lerendegui16} is particularly well suited for these challenging cases. As described in~\cite{Koehler00}, the high peak flux becomes one of the most important features when measuring radioactive samples because it allows for a reduction of the sample-activity background contribution relative to the time-interval where the neutron capture yield is measured.
However, in order to exploit the large peak-neutron flux one requires also radiation detectors with a fast time-response and a high count-rate capability. State-of-the-art C$_6$D$_6$ detectors with a volume of $\sim$1~l~\cite{Plag03} have a relatively large efficiency, which in turn requires a large sample-detector distance to avoid excessive signal pile-up and dead time arising from the high count-rate conditions of about ~1~MHz, or more. Also, the i-TED system described in the preceding section is presently limited to count rates of about 500~kHz~\cite{petsys}, owing to the ASIC-readout scheme implemented to cope with the 1280 readout channels of the twenty position-sensitive detectors. It is expected that future developments will help to enhance the ASIC event-rate capability and possibly, make the i-TED system also useful for measurements in the high-flux conditions of EAR2.

\begin{figure}[!htbp]
    \centering
    \includegraphics[width=\columnwidth]{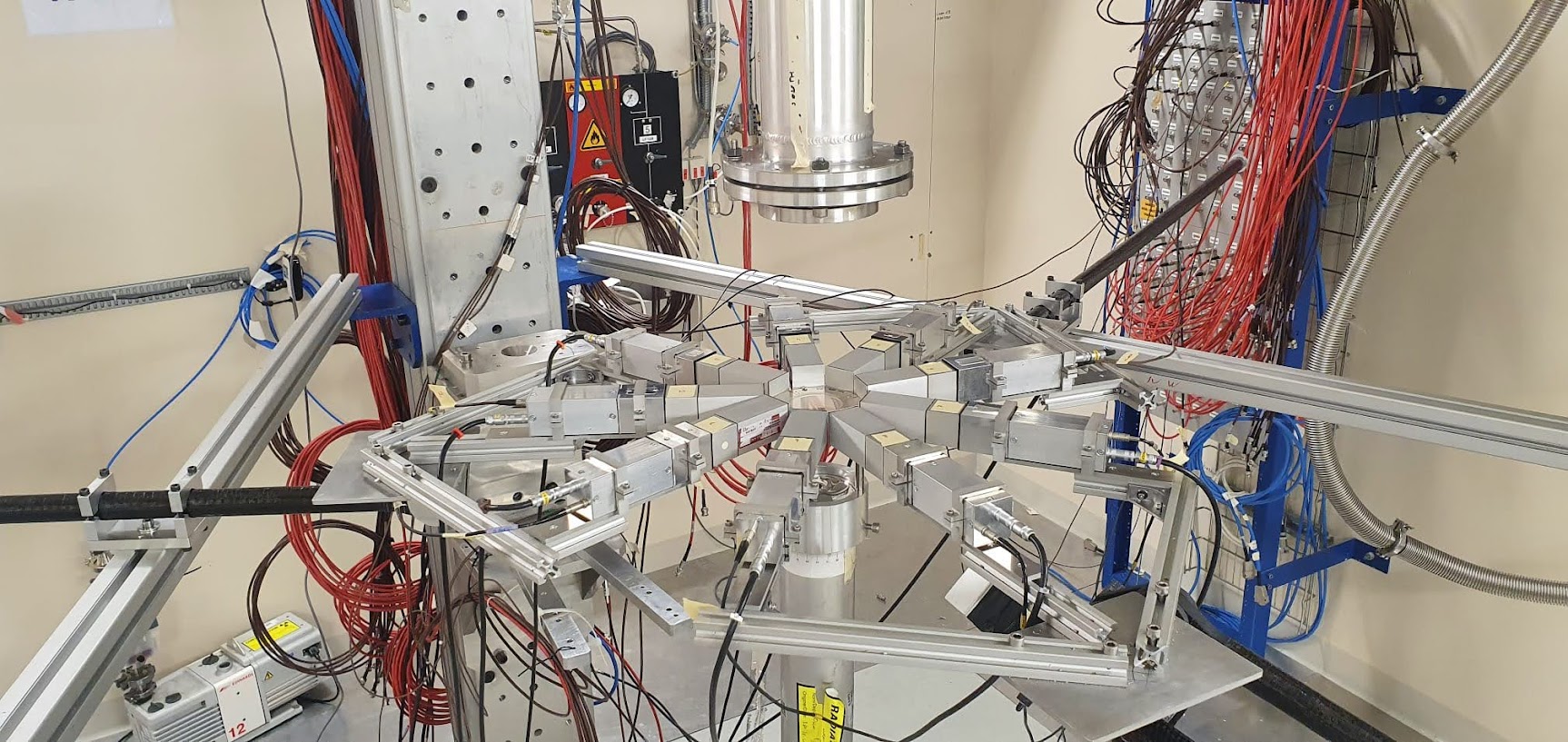}
    \includegraphics[width=\columnwidth]{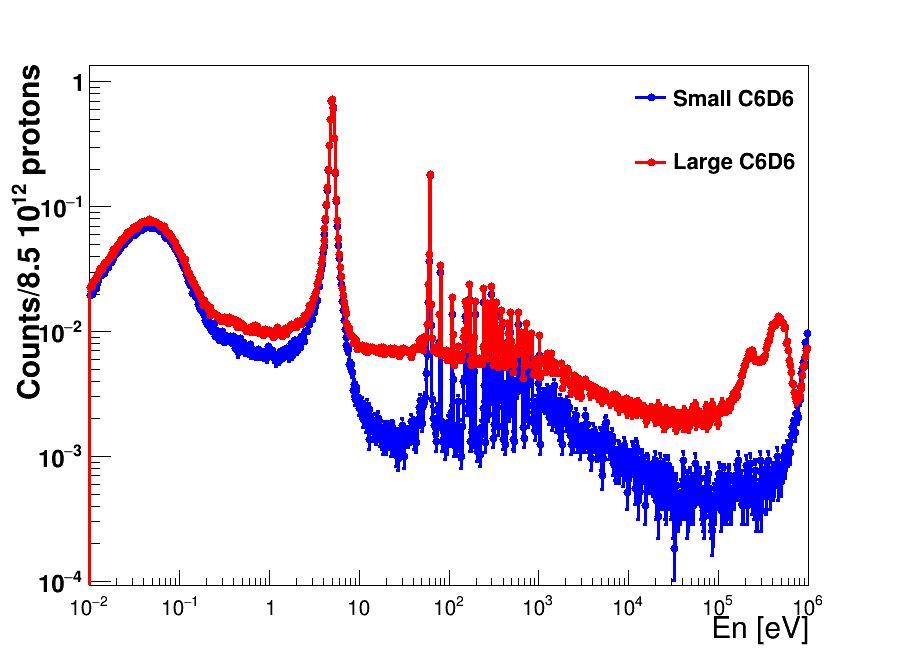}
    \caption{(Top) Photograph of the capture setup based on an array of small-volume C$_6$D$_6$ detectors used for the $^{94}$Nb($n,\gamma$) experiment in 2022 at CERN n\_TOF EAR2. (Bottom) Capture-spectra for $^{197}$Au($n,\gamma$) measured with a conventional large C$_6$D$_6$ detector and with a small volume C$_6$D$_6$ detector in EAR2. Both spectra have been normalized to the peak of the 4.9~eV resonance.}
    \label{fig:sTED}
\end{figure}

To overcome the count-rate limitations of conventional C$_6$D$_6$ detectors an array of nine small-volume (49~ml) C$_6$D$_6$ detectors~\cite{Alcayne22}, was implemented in a compact-ring configuration~\cite{Balibrea22b} around the capture sample in EAR2 as shown in Fig.~\ref{fig:sTED}. 
The main advantage of this innovative setup is that the small detection volume allows one to place the detectors much closer to the capture sample under study, and thus enhance also the efficiency for true capture $\gamma$-rays and increase the signal-to-background ratio (SBR) with respect to previous set-ups based on larger C$_6$D$_6$ detectors placed further away from the beam-line. The improvement in SBR is shown in the bottom panel of Fig.~\ref{fig:sTED}, which shows an enhanced signal-to-background ratio for the $^{197}$Au($n,\gamma$) reaction over most of the energy range when measured with the small-volume C$_6$D$_6$ detectors.
The set-up shown in Fig.~\ref{fig:sTED} was used in the 2022 campaign for the measurement of the $^{94}$Nb($n,\gamma$) cross section~\cite{Balibrea21}. The $^{94}$Nb sample used for this TOF experiment was produced by high-fluence neutron irradiation of hyperpure niobium samples\cite{Moe00} in the V4 tube of the ILL-Grenoble reactor.  The final sample contained a total amount of $9\times10^{18}$~atoms with an activity dominated (10~MBq) by the $\beta$-decay of $^{94}$Nb (2$\times 10^4$~y). 
The results from this experiment are expected to shed light on isotopic anomalies observed in pre-solar SiC grains~\cite{Lugaro03}, which apparently require an unexpectedly large $s$-process contribution to the abundance of $^{94}$Mo.

\section{New astrophysics prospects at NEAR using GEAR and CYCLING}\label{sec:cycling}
%\section{Going exotic with CYCLING}\label{sec:cycling}
The combination of neutron-TOF with activation measurements, when feasible, may deliver complementary and more accurate information on a specific cross section (see Table~II in Ref.~\cite{Kaeppeler11}). When applicable, the activation technique shows an unsurpassed sensitivity for the measurement of minuscule sample quantities, as it has been demonstrated for samples of only $\sim10^{14}$ to $10^{15}$~atoms~\cite{Reifarth03,Uberseder07}.

Following this logic, one of the most recent efforts at n\_TOF concerns the development of the neutron-activation station NEAR~\cite{Ferrari22,Patronis22}, aiming at exploiting the large neutron-fluxes in the proximity of the spallation target. Preliminary MC calculations~\cite{Mengoni20} show the possibility of using suitable filters and moderation materials for producing quasi-Maxwellian neutron-energy spectra over a broad range between a few and several hundreds~keV. A detailed description of the new NEAR installation will be reported in Ref.~\cite{Patronis22} and preliminary flux characterization measurements have been carried out recently~\cite{Stamati22}. Many of the latter measurements have been carried out at the Gamma-ray spectroscopy Experimental ARea (GEAR) of n\_TOF, which is based on a CANBERRA HPGe detector GR5522 supplemented with convenient shielding~\cite{Patronis22}. This station is available for conventional neutron-activation measurements where $\gamma$-rays from the decay of the activation products with half-lives longer than a few hours are measured.

Because of the low duty cycle the average neutron fluence attainable at NEAR is expected to be comparable to the one available in the past at FZK~\cite{Reifarth03} or currently at other activation facilities~\cite{Alzubaidi16,Fernandez20}. However, one of the unique features at NEAR will be the possibility to perform activation measurements on small samples of highly isotopically enriched (or even pure) material, which can be produced in sufficient quantities at the nearby ISOLDE~\cite{Ballof2020} and MEDICIS facilities~\cite{Gadelshin20}. 
%There are many long-lived radioactive isotopes that can be produced in the latter installations and conveniently transported within CERN to the nearby NEAR station for activation experiments of interest, both for astrophysics and nuclear technologies. 
%Because some of these measurements will be described in Ref.~\cite{Patronis22}, i
In addition to the GEAR station, there is another planned station for fast cyclic-activation measurements at NEAR called CYCLING~\cite{Lerendegui22}. The fast-cyclic activation technique was pioneered at FZK-Karslruhe~\cite{Beer94}, where it was applied to measure the neutron-capture cross section of several nuclides of relevance for nucleosynthesis in AGB stars, such as $^{107,109}$Ag($n, \gamma$)~\cite{Beer94}, $^{26}$Mg($n, \gamma$)\cite{Mohr98}, $^{50}$Ti($n, \gamma$)\cite{Sedyshev99} and $^{19}$F($n, \gamma$)\cite{Uberseder07}. It is worth noting that measurements on isotopes with activation products with half-lives as short as $\sim 11$~s ($^{20}$F) are accessible with this technique. The CYCLING station will enable the repetition of a short irradiation, followed by a rapid transport to a detector, where the measurement of the decay will take place and subsequently transported back to the irradiation position. This process is repeated for a number of cycles thus enhancing counting statistics and signal-to-background ratio for short-lived nuclei.

%However, it is worth recalling here that for many $s$-process branching nuclei the activation technique cannot be applied because the daughter nucleus is stable. For example, from the list of 21 relevant $s$-process branching nuclei (Table~III in Ref.~\cite{Kaeppeler11}) only neutron-capture in half of the listed nuclei leads to unstable products and only capture on $^{163}$Ho leads to a product nucleus, $^{163}$Ho, with a half-live shorter than several hours, suitable for the fast-cyclic technique. Indeed,  

Thus, with the future combination of ISOLDE and GEAR-CYCLING it may become possible to access also direct neutron-capture measurements on several unstable nuclei of interest for the study of $s$-process branchings, and also for the more exotic intermediate $i$-process of nucleosynthesis~\cite{Cowan77}. The $i$ process involves neutron capture at neutron densities of $10^{13}–10^{16}$ cm$^{-3}$, in between the $s$ and $r$ processes. Recently, the $i$ process attracted significant interest because it might explain the abundance pattern of a special kind of Carbon-Enhanced Metal-Poor stars (CEMPs), called CEMP-s/r~\cite{Hampel2016}. The site of the $i$-process has been identified as the very late thermal pulse H-ingestion of post-AGB stars. Recent studies show also the relevance of this mechanism for the early generation of stars~\cite{Heger02,Frebel05}.

%At NEAR expected neutron fluences will reach values in the range of 10$^8$-10$^9$~n/cm$^2$/pulse, which are comparable to those that were available in the past at FZK operating the Van de Graaff accelerator at DC proton currents of about 100~$\mu$A~\cite{Kaeppeler11}. Activation measurements have been carried out successfully with radioactive samples of only $\sim$10$^{14}$~atoms~\cite{Reifarth03}-10$^{15}$~atoms~\cite{Uberseder09}. 

%The estimated cool-down and access time to the NEAR station is of at least 4~h and, therefore, cases where the half-lives of the daughter nuclei are shorter than that quantity can be considered unique for the future CYCLING station. 

One case of interest is neutron capture on $^{135}$Cs (t$_{1/2}$ = 2~Myr). The stellar neutron-capture rate of $^{135}$Cs is relevant for the interpretation of the $s$-process branching at $^{134}$Cs (t$_{1/2}$ = 2~yr)~\cite{Kaeppeler11,Patronis04} and also for $i$-process nucleosynthesis, as discussed latter. 

A suitable sample of $^{135}$Cs could be ion-implanted at ISOLDE.  After, characterization and activation at NEAR the decay of the activation product, $^{136}$Cs (t$_{1/2}$=13~d), could be measured at the GEAR station or any other low-background laboratory. The neutron capture of $^{135}$Cs at $kT=25$~keV has been already measured at FZK~\cite{Patronis04} and therefore this measurement could be a good benchmark case for the performance of the new installation. In addition, at NEAR the MACS could be also completed for other neutron energy ranges around  $kT=8$~keV and $kT=90$~keV, where presently there is no experimental information available.

In the high neutron fluxes characteristic of the $i$-process it has been found~\cite{Bertolli2013} that variations in the neutron-capture rates of some specific radioactive isotopes around the $N=82$ neutron-shell closure could affect elemental ratio predictions, involving the benchmark (observable) elements Ba, La and Eu~\cite{Bertolli2013}. Some of the involved reactions, such as $^{137}$Cs($n,\gamma$) may become accessible at NEAR. Commercial samples of $^{137}$Cs (t$_{1/2}$=30~yr) are available and could be used for this measurement. A sample of about $2\times10^{14}$~atoms and an activity of less than 200~kBq (662~keV $\gamma$-rays) could be a suitable option. Capture on $^{137}$Cs leads either directly or via the detour of the shorter-lived $^{138m}$Cs ($t_{1/2}=3$~m) to the activation product $^{138g}$Cs ($t_{1/2}=33$~m) that emits a significant $\gamma$-ray intensity at 1.4 MeV. Owing to the short half-life it could be best measured at the CYCLING station.

As reported in Ref.~\cite{Choplin21}, an AGB star experiencing $s$- or $i$-process nucleosynthesis would show very different isotopic fractions which, although challenging, could be inferred from observations. Thus, several isotopes of Ba, Nd, Sm and Eu may be used as tracers of $i$-process nucleosynthesis. For example, under $i$-process conditions the final abundance of $^{137}$Ba is larger than that of $^{138}$Ba. $^{138}$Ba, with $N=82$, has a very small neutron-capture cross section, acting as a bottleneck and therefore being copiously produced by the $s$ process. The relatively large $i$-process abundance of $^{137}$Ba is due to the decay of $^{137}$Cs which, at variance with the $s$ process, can be easily reached in $i$-process conditions. Therefore, the aforementioned $^{135}$Cs($n,\gamma$) and $^{137}$Cs($n,\gamma$) cross section measurements could provide a valuable input information for $i$-process models and observations. In addition, the measurement of the intermediate $^{136}$Cs($n,\gamma$) may become feasible, assuming that a sample with sufficient mass could be produced at ISOLDE and later activated at NEAR. After the neutron activation and sufficient waiting time to let the $^{136}$Cs (t$_{1/2}$=13~d) in the sample decay, one could measure the activity of the activation product $^{137}$Cs (t$_{1/2}$=30~yr) at the GEAR station.
Other similar cases related to the $i$-process tracers discussed in Ref.~\cite{Choplin21} might be also accessible at NEAR, such as neutron capture on $^{144}$Ce (t$_{1/2}$=285~d) leading to $^{145}$Ce (t$_{1/2}$=3~m). However, the feasibility with CYCLING needs to be studied in detail owing to the $\gamma$-ray activity from neighbouring decays (mainly $^{144}$Pr).

Finally, there are many other neutron-capture reactions of interest for the $i$-process, such as neutron capture on $^{66}$Ni (t$_{1/2}$=55~h), which represents one of the major bottle-necks in $i$-process models~\cite{McKay2020} or neutron capture on $^{72}$Zn (t$_{1/2}$=46~h) that determines the $i$-process abundance of Ge~\cite{McKay2020}. However, in these cases activation measurements become prohibitive due to the large sample activity, which typically exceeds 100~MBq for sample quantities of about 10$^{12-13}$~atoms. For this reason, indirect methods such as surrogate reactions using storage rings~\cite{Perez20,Jurado21}, may represent the most promising alternative to obtain experimental information and to constrain the stellar environments.

\section{Summary and outlook}\label{sec:summary}
This article has presented a few technical contributions of n\_TOF to the field of neutron-capture experiments of astrophysical interest. These works have been key, on the one hand, to address the accuracy of the measurements, and even enhance the systematic precision for this type of studies~\cite{Tain04}, an aspect which is closely connected with the 4-5\% systematic error commonly required for reliable astrophysical interpretation of observational data or meteorites analysis~\cite{Kaeppeler11,Cescutti19,Nishimura18b}.
Although historically, a large effort has been invested in reducing the intrinsic neutron-sensitivity of the detection apparatus, detailed MC calculations~\cite{Zugec14} showed that, in many situations, the background level is dominated by scattered neutrons, which are captured in the surroundings of the detectors, rather than in the detection system itself. In this respect, a novel i-TED detection system~\cite{Domingo16} based on $\gamma$-ray imaging has been developed, which allows one to attain a significant improvement in signal-to-background ratio for such specific cases in the keV-energy range of astrophysical interest~\cite{Babiano21}. This system has been employed at CERN n\_TOF for the first measurement of the $^{79}$Se($n,\gamma$) cross section, which is one of the main branching points in the weak $s$ process~\cite{Kaeppeler11}. Further, for the measurement of highly-radioactive samples, such as the one of $^{94}$Nb described in Sec.~\ref{sec:sTED}, a new array of very small-volume C$_6$D$_6$ detectors was developed and implemented, which enabled also for a significant improvement in terms of signal-to-background ratio with respect to state-of-the-art C$_6$D$_6$ detectors. This measurement, carried out also in 2022 at CERN n\_TOF, will help to shed light on isotopic Mo-anomalies observed in pre-solar SiC grains~\cite{Lugaro03}.
Future ideas and proposals at n\_TOF are related to the new NEAR experimental area for exploiting also the neutron-activation technique in measurements of astrophysical interest. In this respect, current efforts to design a station for fast cyclic activation measurements (CYCLING) have been also presented. This installation could help to access for the first time to direct neutron-capture cross sections on radioactive isotopes which are of great interest for the intermediate process of nucleosynthesis.

\section*{Acknowledgment}
For the purpose of open access, the author has applied
a Creative Commons Attribution (CC BY) licence to any
Author Accepted Manuscript version arising from this sub-
mission.
Part of this work has been carried out in the framework of a project funded by the European Research Council (ERC) under the European Union's Horizon 2020 research and innovation programme (ERC Consolidator Grant project HYMNS, with grant agreement No.~681740). The authors acknowledge support from the Spanish Ministerio de Ciencia e Innovaci\'on under grants PID2019-104714GB-C21, FPA2017-83946-C2-1-P, FIS2015-71688-ERC, FPA2016-77689-C2-1-R, RTI2018-098117-B-C21, CSIC for funding PIE-201750I26, European H2020-847552 (SANDA) and by funding agencies of participating institutes. 

This article belongs to a series of articles devoted to the memory of Franz K\"appeler. The present work contains some of the developments where he was involved or witnessed and, some other contributions which came up more recently.  In any case, all of them have undoubtedly benefited from the motivation and creativity that Franz inspired in all of us. Thank you Franz.

\bibliography{bibliography}

\end{document}